\documentclass[11pt]{article}
\usepackage[russian]{babel}
\usepackage{amssymb,amsmath,amsthm}
\usepackage{graphicx}
\usepackage{epstopdf}
\newcommand{\bc}{\begin{center}}
\newcommand{\ec}{\end{center}}
\newcommand{\be}{\begin{equation}}
\newcommand{\ee}{\end{equation}}
\newcommand{\bea}{\begin{eqnarray}}
\newcommand{\eea}{\end{eqnarray}}
\newcommand{\ba}{\begin{array}}
\newcommand{\ea}{\end{array}}
\newcommand{\edc}{\end{document}}

\begin{document}
УДК 517.98
\begin{center}
\textbf{\Large {Крайность единственной трансляционно-инвариантной меры Гиббса для НС моделей на дереве Кэли порядка три }}\\
\end{center}

\begin{center}
Р.М.Хакимов\footnote{Наманганское отделение института Математики имени В.И.Романовского АН РУз, Наманган, Узбекистан.\\
E-mail: rustam-7102@rambler.ru}, К.О.Умирзакова\footnote{Наманганский государственный университет, Наманган, Узбекистан..\\
E-mail: kamola-0983@mail.ru}
\end{center}

\begin{abstract} Данная работа посвящена изучению плодородных Hard-Core (HC) моделей
с тремя состояниями и параметром активности $\lambda>0$ на дереве Кэли порядка три.
Известно, что существуют четыре типа таких моделей. Для двух из них найдены области
(не) крайности единственной трансляционно-инвариантной меры Гиббса. Кроме того, для
одной из рассматриваемых моделей найдены условия, при которых крайняя мера не единственна.
\end{abstract}

\textbf{Ключевые слова}: дерево Кэли, конфигурация, HC-модель, плодородный граф, 
мера Гиббса, трансляционно-инвариантные меры, крайность меры.

\begin{center}
\textbf{1. ВВЕДЕНИЕ}
\end{center}

В теории мер Гиббса основной задачей, с одной стороны, является полное описание
всех предельных мер Гиббса для данного гамильтониана. Каждой предельной мере
Гиббса сопоставляется одна фаза физической системы, и если мера Гиббса не единственна,
то говорят, что существует фазавый переход, т.е. физическая система меняет свое состояние
при изменении температуры. С другой стороны, так как множество всех предельных мер Гиббса
образует выпуклое компактное подмножество в множестве всех вероятностных мер
(см. \cite {6}-\cite {R}) и каждая точка этого выпуклого множества однозначно
разлагается по его крайним точкам, то особый интерес представляет описание всех
крайних точек этого выпуклого множества, т. е. крайних мер Гиббса.

Надо отметить, что несмотря на многочисленные работы, посвященных изучению мер Гиббса,
ни для одной модели не было получено полное описание всех предельных мер Гиббса. В работе
\cite{KRK} дано полное описание трансляционно-инвариантных мер Гиббса (ТИМГ) для
ферромагнитной модели Поттса с $q$-состояниями и показано, что их максимальное количество
равно $2^q-1$.

НС-модель (жесткий диск, жесткая сердцевина) на $d$-мерной решетке $ \mathbb Z ^ d$
была введена и изучена в работе \cite {Maz}. В работе \cite{bw} выделены плодородные
HC модели, соответствующие графам "петля"\,, "свисток"\,, "жезл"\, и "ключ".
Работы \cite{7}-\cite {RKh1} посвящены изучению мер Гиббса для HC-моделей
с тремя состояниями на дереве Кэли порядка $k\geq1$. В частности, в работе
\cite{7} изучена HC-модель на дереве Кэли и доказано, что ТИМГ для
этой модели единственна. Кроме того, при некоторых условиях на параметры
доказана неединственность периодических мер Гиббса с периодом два. В работе
\cite{MRS} были изучены трансляционно-инвариантные и периодические меры Гиббса
для HC модели в случае "ключ"\, на дереве Кэли и была доказана единственность
ТИМГ для любой положительной активности $\lambda$. В работах \cite{Ro} и \cite {XR1}
в случаях "петля"\, и "жезл"\ дано полное описание ТИМГ на дереве Кэли порядка два и три,
соответственно, а в работе \cite {RKh1} в этих случаях доказано существование не менее
трех ТИМГ на дереве Кэли произвольного порядка. Кроме того, в \cite {RKh1} найдены
области (не) крайности ТИМГ на дереве Кэли порядка $k=2$. А работа \cite {KKR} посвящена
полному описанию трансляционно-инвариантных и изучению периодических мер Гиббса для НС моделей
с тремя состояниями с внешним полем. В работе \cite {KhU} для НС-модели в случае "жезл"\ изучены
существование и крайность два-периодических мер Гиббса. Для ознакомления с другими свойствами
НС-модели (и их обобщениями) на дереве Кэли см. Главу 7 монографии \cite {R}.

В настоящей работе рассматривается плодородные HC-модели с тремя состояниями в случаях
"петля"\, и "жезл"\ на дереве Кэли порядка три. В каждом из этих случаев найден явный вид единственной
ТИМГ и определены области крайности и не крайности этой меры. Кроме того,
в случае "жезл"\ найдены условия, при которых крайняя мера не единственна на дереве Кэли третьего порядка.

\begin{center}
\textbf{2. ПРЕДВАРИТЕЛЬНЫЕ СВЕДЕНИЯ}
\end{center}

Дерево Кэли $\Im^k$ порядка $ k\geq 1 $ - бесконечное дерево, т.е.
граф без циклов, из каждой вершины которого выходит ровно $k+1$
ребер. Пусть $\Im^k=(V,L,i)$, где $V$ есть множество вершин
$\Im^k$, $L-$ множество его ребер и $i-$ функция инцидентности,
сопоставляющая каждому ребру $l\in L$ его концевые точки $x, y \in
V$. Если $i (l) = \{ x, y \} $, то $x$ и $y$ называются  {\it
ближайшими соседями вершины} и обозначаются через $l = \langle x,
y \rangle $.

Для фиксированной $x^0\in V$ положим
$$W_n=\{x\in V\,| \, d(x,x^0)=n\}, \qquad V_n=\bigcup_{m=0}^n W_m, \qquad L_n=\{\langle x,y\rangle\in L| \, x,y\in V_n\},$$
где $d(x,y)$ есть расстояние между вершинами $x$ и $y$ на дереве Кэли, т.е.
количество ребер кратчайшей пути, соединяющей вершины $x$ и $y$.
Будем писать $x\prec y$, если путь от $x^0$ до $y$ проходит через $x$.
Вершину $y$ назовем прямым потомком вершины $x$, если $y\succ x$ и $x,y$
являются ближайшими соседями. Заметим, что в $\Im^k$ всякая вершина $x\neq x^0$
имеет $k$ прямых потомков, а вершина $x^0$ имеет $k+1$ потомков. Множество прямых
потомков вершины $x$ обозначим через $S(x)$, т.е. если $x\in W_n$, то
$$S(x)=\{y_i\in W_{n+1} |  d(x,y_i)=1, i=1,2,\ldots, k \}.$$

\emph{НС модель.} Пусть $\Phi = \{0,1,2\}$ и $\sigma\in \Omega=\Phi^V$ есть конфигурация,
т.е. $\sigma=\{\sigma(x)\in \Phi: x\in V\}$. Иными словами, в этой модели каждой вершине
$x$ ставится в соответствие одно из значений $\sigma (x)\in \Phi=\{0,1,2\}$.
Значения $\sigma (x)=1,2$ означают, что вершина $x$ `занята', а
значение $\sigma (x)=0$ означает, что вершина $x$ `вакантна'.
Множество всех конфигураций на $V$ ($V_n$, $W_n$) обозначается через $\Omega$ ($\Omega_{V_n}$,
$\Omega_{W_n}$).

Рассмотрим множество $\Phi$ как множество вершин некоторого графа
$G$. С помощью графа $G$ определим $G$-допустимую конфигурацию
следующим образом. Конфигурация $\sigma$ называется
$G$-\textit{допустимой конфигурацией} на дереве Кэли (в $V_n$ или
$W_n$), если $\{\sigma (x),\sigma (y)\}$-ребро графа $G$ для любой
ближайшей пары соседей $x,y$ из $V$ (из $V_n$). Обозначим
множество $G$-допустимых конфигураций через $\Omega^G$
($\Omega_{V_n}^G$).

Множество активности \cite{bw} для графа $G$ есть функция $\lambda:G
\to R_+$. Значение $\lambda_i$ функции $\lambda$ в вершине
$i\in\{0,1,2\}$ называется ее ``активностью''.

Для данных $G$ и $\lambda$ определим гамильтониан $G-$HC-модели как
 $$H^{\lambda}_{G}(\sigma)=\left\{%
\begin{array}{ll}
     \sum\limits_{x\in{V}}{\log \lambda_{\sigma(x)},} \ \ \ $ если $ \sigma \in\Omega^{G} $,$ \\
   +\infty ,\ \ \ \ \ \ \ \ \ \  \ \ \ $  \ если $ \sigma \ \notin \Omega^{G} $.$ \\
\end{array}%
\right. $$

Объединение конфигураций $\sigma_{n-1}\in\Phi ^ {V_{n-1}}$ и $\omega_n\in\Phi
^ {W_{n}}$ определяется следующей формулой (см. \cite{GFM})
$$
\sigma_{n-1}\vee\omega_n=\{\{\sigma_{n-1}(x), x\in V_{n-1}\},
\{\omega_n(y), y\in W_n\}\}.
$$

Пусть $\mathbf{B}$ есть $\sigma$-алгебра, порожденная
цилиндрическими подмножествами $\Omega.$ Для любого $n$ обозначим
через $\mathbf{B}_{V_n}=\{\sigma\in\Omega:
\sigma|_{V_n}=\sigma_n\}$ подалгебру $\mathbf{B},$ где
$\sigma|_{V_n}-$ сужение $\sigma$ на $V_n,$ $\sigma_n: x\in V_n
\mapsto \sigma_n(x)-$ допустимая конфигурация в $V_n.$

\textbf{Определение 1}. Для $\lambda >0$ НС-мера Гиббса есть
вероятностная мера $\mu$ на $(\Omega , \textbf{B})$ такая, что для
любого $n$ и $\sigma_n\in \Omega_{V_n}$
$$
\mu \{\sigma \in \Omega:\sigma|_{V_n}=\sigma_n\}=
\int_{\Omega}\mu(d\omega)P_n(\sigma_n|\omega_{W_{n+1}}),
$$
где
$$
P_n(\sigma_n|\omega_{W_{n+1}})=\frac{e^{-H(\sigma_n)}}{Z_{n}
(\lambda ; \omega |_{W_{n+1}})}\textbf{1}(\sigma_n \vee \omega
|_{W_{n+1}}\in\Omega_{V_{n+1}}).
$$

Здесь $Z_n(\lambda ; \omega|_{W_{n+1}})-$ нормировочный множитель с
граничным условием $\omega|_{W_n}$:
$$
Z_n (\lambda ; \omega|_{W_{n+1}})=\sum_{\widetilde{\sigma}_n \in
\Omega_{V_n}}
e^{-H(\widetilde{\sigma}_n)}\textbf{1}(\widetilde{\sigma}_n\vee
\omega|_{W_{n+1}}\in \Omega_{V_{n+1}}).
$$

\textbf{Определение 2.}(\cite{bw}) Граф называется плодородным,
если существует набор активности $\lambda$ такой, что
соответствующий гамильтониан имеет не менее двух ТИМГ.

В этой работе мы рассмотрим случай $\lambda_0=1, \
\lambda_1=\lambda_2=\lambda \ $ и изучим ТИМГ в случаях плодородных графов $G=\textit{петля}$ и $G=\textit{жезл}$:
\[
\begin{array}{ll}
\mbox{\it петля}: &  \{0,0\}\{0,1\}\{0,2\}\{1,1\}\{2,2\};\\
\mbox{\it жезл}: &  \{0,1\}\{0,2\}\{1,1\}\{2,2\}.\\
\end{array} \]

Для $\sigma_n\in\Omega_{V_n}^G$ положим
$$\#\sigma_n=\sum\limits_{x\in V_n}{\mathbf 1}(\sigma_n(x)\geq 1)$$
число занятых вершин в $\sigma_n$.

Пусть $z:\;x\mapsto z_x=(z_{0,x}, z_{1,x}, z_{2,x}) \in R^3_+$
векторнозначная функция на $V$. Для $n=1,2,\ldots$ и $\lambda>0$
рассмотрим вероятностную меру $\mu^{(n)}$ на $\Omega_{V_n}^G$,
определяемую как
\begin{equation}\label{rus2.1}
\mu^{(n)}(\sigma_n)=\frac{1}{Z_n}\lambda^{\#\sigma_n} \prod_{x\in
W_n}z_{\sigma(x),x}.
\end{equation}

Здесь $Z_n-$ нормирующий делитель:
$$
Z_n=\sum_{{\widetilde\sigma}_n\in\Omega^H_{V_n}}
\lambda^{\#{\widetilde\sigma}_n}\prod_{x\in W_n}
z_{{\widetilde\sigma}(x),x}.
$$

Говорят, что последовательность вероятностных мер $\mu^{(n)}$ является
согласованной, если $\forall$ $n\geq 1$ и
$\sigma_{n-1}\in\Omega^G_{V_{n-1}}$:

\begin{equation}\label{rus2.2}
\sum_{\omega_n\in\Omega_{W_n}}
\mu^{(n)}(\sigma_{n-1}\vee\omega_n){\mathbf 1}(
\sigma_{n-1}\vee\omega_n\in\Omega^G_{V_n})=
\mu^{(n-1)}(\sigma_{n-1}).
\end{equation}

\textbf{Определение 3.} Мера $\mu$, определенная формулой
(\ref{rus2.1}) с условием согласованности (\ref{rus2.2}), называется
($G$-)HC-\textit{мерой Гиббса} с $\lambda>0$,
\textit{соответствующей функции} $z:\,x\in V
\setminus\{x^0\}\mapsto z_x$.

Пусть $L(G)-$ множество ребер графа $G$, обозначим через $A\equiv
A^G=\big(a_{ij}\big)_{i,j=0,1,2}$ матрицу смежности $G$, т.е.
$$ a_{ij}\equiv a^G_{ij}=\left\{\begin{array}{ll}
1,\ \ \mbox{если}\ \ \{i,j\}\in L(G),\\
0, \ \ \mbox{если} \ \  \{i,j\}\notin L(G).
\end{array}\right.$$

В следующей теореме сформулировано условие на $z_x$, гарантирующее
согласованность меры $\mu^{(n)}$.

\textbf{Теорема 1.}\cite{Ro} Вероятностные меры
$\mu^{(n)}$, $n=1,2,\ldots$, заданные формулой (\ref{rus2.1}),
согласованны тогда и только тогда, когда для любого $x\in V$ имеют
место следующие равенства:
\begin{equation}\label{rus2.3}\begin{array}{llllll}
z'_{1,x}=\lambda \prod_{y\in S(x)}{a_{10}+
a_{11}z'_{1,y}+a_{12}z'_{2,y}\over
a_{00}+a_{01}z'_{1,y}+a_{02}z'_{2,y}},\\[4mm]
z'_{2,x}=\lambda \prod_{y\in S(x)}{a_{20}+
a_{21}z'_{1,y}+a_{22}z'_{2,y}\over
a_{00}+a_{01}z'_{1,y}+a_{02}z'_{2,y}},
\end{array}
\end{equation}
где $z'_{i,x}=\lambda z_{i,x}/z_{0,x}, \ \ i=1,2$.\

В (\ref{rus2.3}) мы полагаем, что $z_{0,x} \equiv1$ и $z_{i,x}=z'_{i,x}>0, i=1,2.$ Тогда в силу теоремы 1
существует единственная $G$-HC-мера Гиббса $\mu$ тогда и только тогда, когда для любых функций
$z:x\in V \longmapsto z_{x}=(z_{1,x},z_{2,x})$ выполняется равенство
\begin{equation}\label{rus2.4}
z_{i,x}=\lambda\prod_{y\in S{x}}\frac{a_{i0}+a_{i1}z_{1,y}+a_{i2}z_{2,y}}{a_{00}+a_{01}z_{1,y}+a_{02}z_{2,y}},    i=1,2.
\end{equation}
Рассмотрим трансляционно-инвариантные решения, в которых $z_x=z\in
R^2_+$, $x\neq x_0$.

\begin{center}
\textbf{3. КРАЙНОСТЬ ТРАНСЛЯЦИОННО-ИНВАРИАНТНЫХ МЕР ГИББСА}
\end{center}

\textbf{Случай $G=\textit{петля}$}. В этом случае предполагая $z_x=z$, из (\ref{rus2.4}) получим
следующую систему уравнений:

\begin{equation}\label{rus3.1} \left\{\begin{array}{ll}
z_1=\lambda\left({1+ z_1\over 1+z_1+z_2}\right)^k,\\[2mm]
z_2=\lambda\left({1+z_2\over 1+z_1+z_2}\right)^k.
\end{array}\right.
\end{equation}

Для ТИМГ в случае $G=\textit{петля}$ известны следующие факты:

\begin{itemize}
\item[$\bullet$] В случае $k=2$ ($k=3$) доказано, что при $\lambda\leq \frac{9}{4}$ ($\lambda\leq\frac{32}{27}$)
существует ровно одна ТИМГ $\mu_0$, при $\lambda>\frac{9}{4}$ ($\lambda>\frac{32}{27}$) существуют ровно
три ТИМГ $\mu_0, \mu_1, \mu_2$ (см. \cite{Ro}, \cite{XR1}).

\item[$\bullet$] В случае $k>3$ доказано, что при $\lambda\leq\lambda_{cr}$ существует ровно
одна ТИМГ, при $\lambda>\lambda_{cr}$ существуют не менее трех ТИМГ, где
$\lambda_{cr}={1\over k-1}\cdot\left({k+1\over k}\right)^k$ (см. \cite{RKh1}).

\item[$\bullet$] В случае $k=2$ показано, что мера $\mu_0$ при $0<\lambda<\lambda_0$ и меры
$\mu_1, \mu_2$ при $\frac{9}{4}<\lambda<{1\over 2}(5\sqrt{2}-1)$ являются крайними и мера $\mu_0$ при $\lambda>\lambda_0$
не является крайней, где $\lambda_0\approx 7.0355$ (см. \cite{RKh1}).
\end{itemize}

Рассмотрим (\ref{rus3.1}) при $k=3$:

\begin{equation}\label{rus3.2} \left\{\begin{array}{ll}
z_1=\lambda\left({1+ z_1\over 1+z_1+z_2}\right)^3,\\[2mm]
z_2=\lambda\left({1+z_2\over 1+z_1+z_2}\right)^3.
\end{array}\right.
\end{equation}
Известно \cite{XR1}, что единственной ТИМГ $\mu_0$ соответствует единственное решение уравнения
$$z=\lambda\left(\frac{1+z}{1+2z}\right)^3,$$
которое получается из (\ref{rus3.2}) при $z_1=z_2=z$.
После обозначений $\sqrt[3]{z}=x,  \sqrt[3]{\lambda}=a$, из последнего уравнения получим уравнение
\begin{equation}\label{rus3.5}
2x^4-ax^3+x-a=0,
\end{equation}
которого решим по методу Феррари из линейной алгебры. Перепишем последнее уравнение
$$x^4-\frac{a}{2}x^3+\frac{x}{2}-\frac{a}{2}=0$$
и сделаем замену $x=y+\frac{a}{8}$. Тогда получим
$$y^4-\frac{3}{32}a^2y^2+\left(\frac{1}{2}-\frac{a^3}{64}\right)y-\frac{3a^4}{4096}-\frac{7}{16}a=0.$$
В последнем уравнении, введя новый вспомогательный параметр $t$, тождественно преобразуем его. Тогда будем иметь
следующее уравнение:
$$\left(y^2-\frac{3}{64}a^2+t\right)^2-\left[2ty^2-\left(\frac{1}{2}-\frac{a^3}{64}\right)y-
\frac{3a^2}{32}t+\frac{3a^4}{1024}+\frac{7a}{16}+t^2\right]=0.$$
Далее, параметр $t$ выберем так, чтобы многочлен в квадратных скобках был полным квадратом,
т.е. его дискриминант
$$D=\left(\frac{1}{2}-\frac{a^3}{64}\right)^2-8t\left(\frac{3}{1024}a^4+\frac{7}{16}a+t^2-\frac{3}{32}a^2t\right)=0$$
или
$$8t^3-\frac{3}{4}a^2t^2+\left(\frac{3}{128}a^4+\frac{7}{2}a\right)t-\left(\frac{1}{2}-\frac{a^3}{64}\right)^2=0.$$
Решим его с помощью формулы Кардано. Тогда получим
$$t_0(a)=\frac{1}{24}\sqrt[3]{-108a^3+216+12\sqrt{81a^6+3792a^3+324}}-$$
$$-\frac{7}{2}\frac{a}{\sqrt[3]{-108a^3+216+12\sqrt{81a^6+3792a^3+324}}}+\frac{1}{32}a^2.$$
Отсюда при $t=t_0$ и заметив, что $x=y+\frac{a}{8}$, для решений (\ref{rus3.5}) будем иметь:
$$x_{1,2}=\frac{1}{2}\sqrt{2t_0(a)}\pm\frac{1}{2}\sqrt{-2t_0(a)+\sqrt{\frac{2}{t_0(a)}}\left(\frac{1}{2}-\frac{a^3}{64}\right)+\frac{3a^2}{16}}+\frac{a}{8},$$
$$x_{3,4}=-\frac{1}{2}\sqrt{2t_0(a)}\pm\frac{1}{2}\sqrt{-2t_0(a)-\sqrt{\frac{2}{t_0(a)}}\left(\frac{1}{2}-\frac{a^3}{64}\right)+\frac{3a^2}{16}}+\frac{a}{8}.$$
С помощью компьютерного анализа для единственного положительного решения уравнения (\ref{rus3.5}) будем иметь следующее:
при $0<a<a_0$ оно имеет следующий вид:
\begin{equation}\label{rus3.3}
x_1=\frac{1}{2}\sqrt{2t_0(a)}+\frac{1}{2}\sqrt{-2t_0(a)+\sqrt{\frac{2}{t_0(a)}}\left(\frac{1}{2}-\frac{a^3}{64}\right)+\frac{3a^2}{16}}+\frac{a}{8},
\end{equation}
а при $a>a_0$:
\begin{equation}\label{rus3.4}
x_3=-\frac{1}{2}\sqrt{2t_0(a)}+\frac{1}{2}\sqrt{-2t_0(a)-\sqrt{\frac{2}{t_0(a)}}\left(\frac{1}{2}-\frac{a^3}{64}\right)+\frac{3a^2}{16}}+\frac{a}{8},
\end{equation}
где
$$a_0=\frac{1}{29988}\sqrt[3]{(-405+51\sqrt{177})^5}+\frac{331}{4998}\sqrt[3]{(-405+51\sqrt{177})^2}\approx3.174802104.$$

Таким образом, справедливо следующее утверждение.

\textbf{Утверждение 1.} Единственное положительное решение $(z,z)$ системы уравнений (\ref{rus3.2}) при $\lambda>0$ имеет следующий вид:
\begin{equation}\label{rus4.1}(z,z)=\left\{%
\begin{array}{ll}
    (x_1^3,x_1^3), \ \ \ $ если $ 0<\lambda<\lambda^{*}$,$ \\
    (x^3,x^3) ,\ \ \  $   если $ \lambda=\lambda^{*} $, $ \\
    (x_3^3,x_3^3) ,\ \ \  $   если $ \lambda>\lambda^{*} $, $ \\
\end{array}%
\right.\end{equation}
где $a=\sqrt[3]{\lambda}$, $\lambda^{*}=a^3_0\approx32$, $x=x_1=x_3$ и $x_1, \ x_3$ определены
формулами (\ref{rus3.3}), (\ref{rus3.4}), соответственно.

Итак, мы имеем ТИМГ $\mu_0$, соответствующую этому решению.
Чтобы изучить (не) крайность меры $\mu_0$ воспользуемся методами
из работ \cite{RKh1}, \cite{MSW}, \cite{KS}, \cite{KR} и \cite{Kr1}
для ТИМГ. Найдем условия не крайности меры $\mu_0$.

Известно \cite{RKh1}, что матрица вероятностных переходов для этой меры имеет вид ($z_1=z_2=z$):
$$
\mathbb P=\left(%
\begin{array}{cccccc}
 {1\over 1+2z} &  {z\over 1+2z} &  {z\over 1+2z} \\[3 mm]
  {1\over 1+z} & {z\over 1+z} & 0 \\[3 mm]
  {1\over 1+z} & 0 & {z\over 1+z} \\
  \end{array}%
\right)
$$
и ее собственные значения:
$$s_1=-{1\over (z+1)(2z+1)}, \ \ s_2={z\over z+1}, \ s_3=1.$$
Отсюда
$$s_0=max\{|s_1|, |s_2|\}=\left\{%
\begin{array}{ll}
    |s_1|, \ \ \ $ если $ {z<{1\over 2}} $,$ \\
   |s_2| ,\ \ \  $   если $ {z>{1\over 2}} $, $ \\
\end{array}%
\right.$$
где $z$ имеет вид (\ref{rus4.1}).

\textit{Не крайность меры $\mu_0$.} Для определения области не крайности меры $\mu_0$ проверим условие Кестена-Стигума
$ks_0^2>1$, где $s_0-$ второе максимальное по абсолютной величине собственное значение матрицы $\mathbb P$.

Так как выражение для $z$ очень громоздкое, восьползуемся компьютерным анализом. С помощью программы MapLE
можно увидеть, что $h(z)=z-\frac{1}{2}>0$ при любых $\lambda>0$ (см. Рис.1). Значит, условие не крайности имеет вид
$3s_2^2=3\left({z\over z+1}\right)^2>1$ или $g(z)=(\sqrt{3}-1)z-1>0.$ Из этого неравенства следует, что оно верно при $\lambda>\hat{\lambda}\approx0.8094705632$, т.е. мера $\mu_0$ не является крайней при этом условии (см. Рис.2).

\begin{center}
\includegraphics[width=6cm]{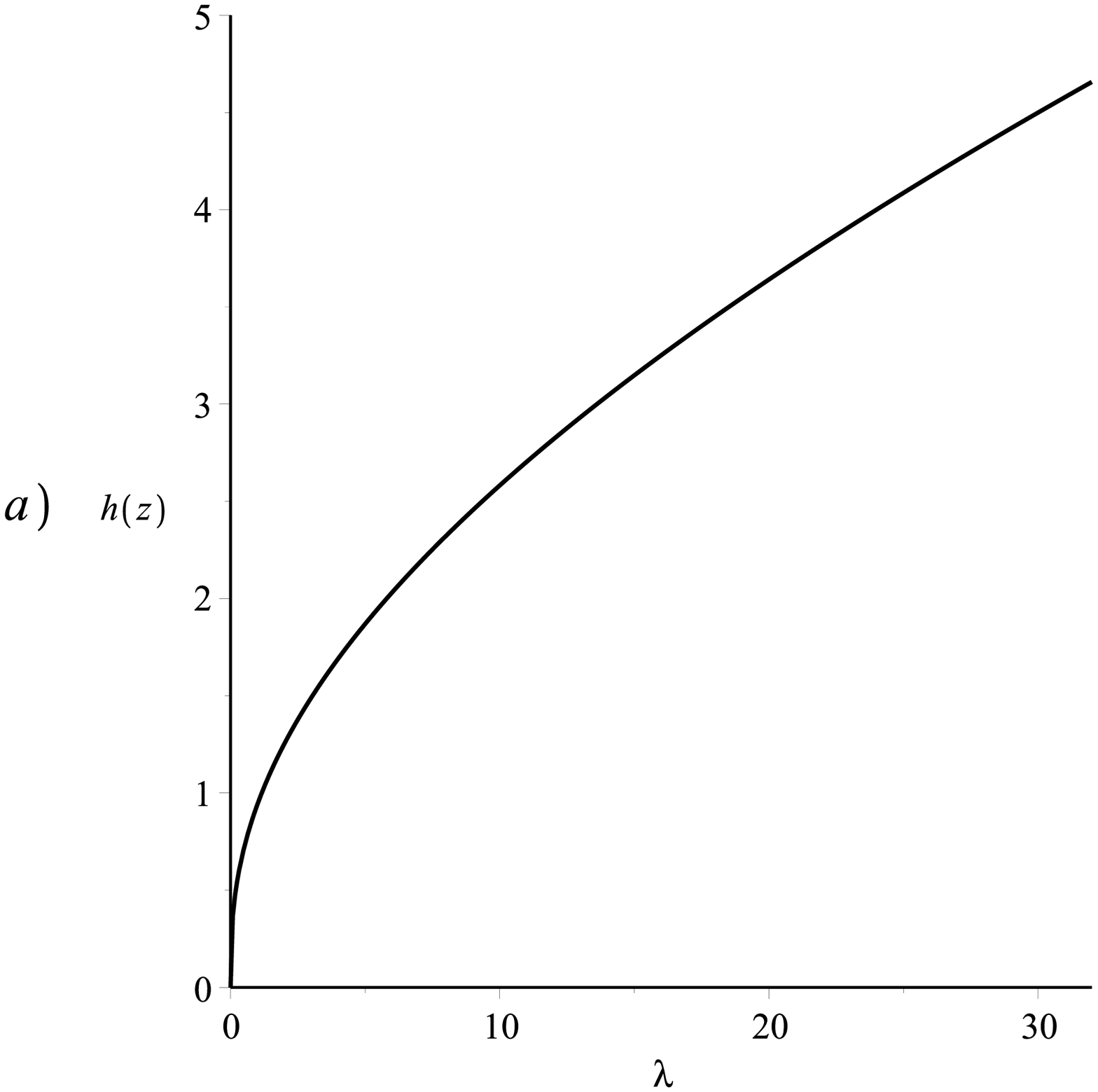} \  \includegraphics[width=6cm]{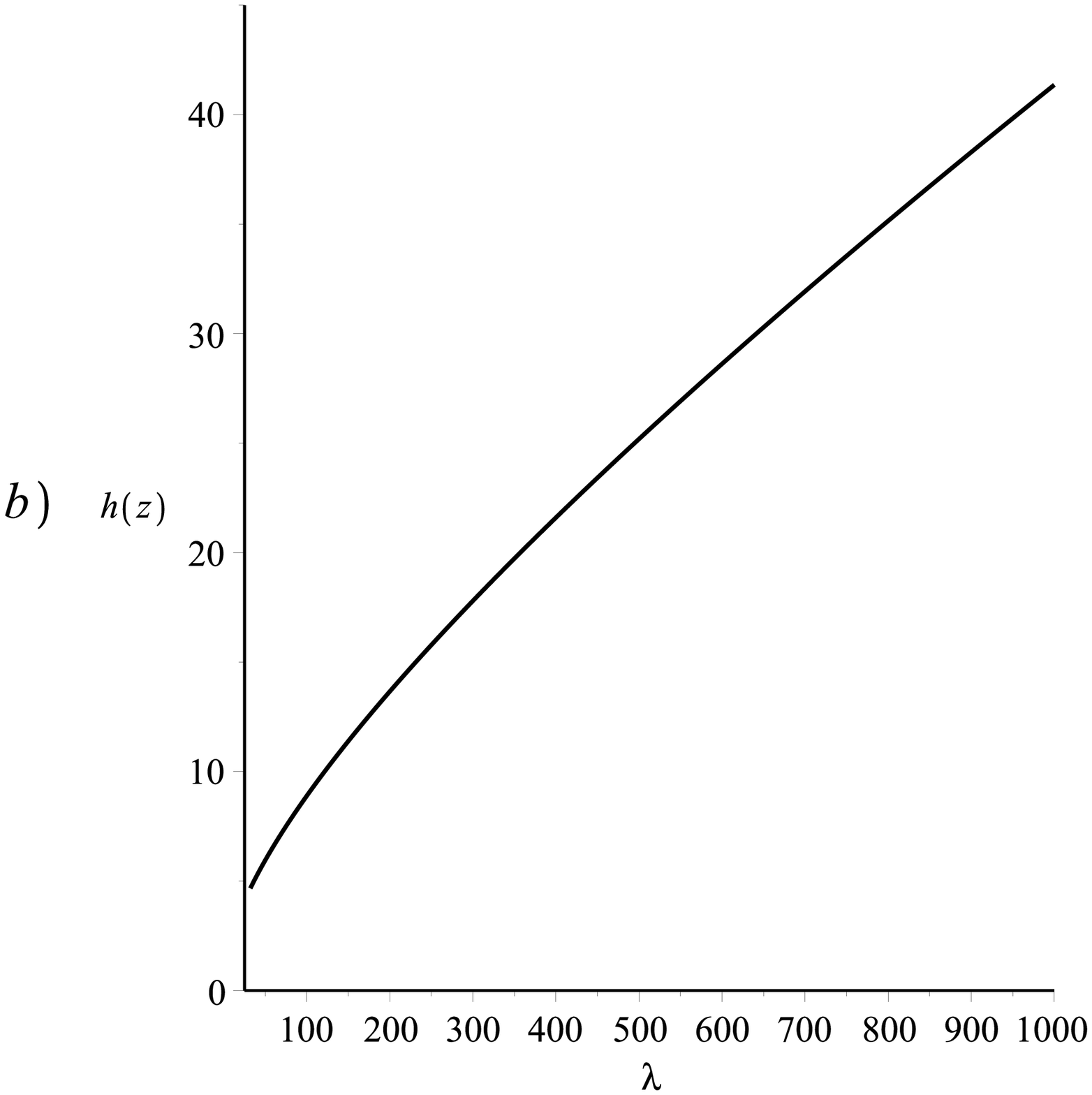}
\end{center}
\begin{center}{\footnotesize \noindent
 Рис.~1. a) График функции $h(z)$ при $z=x_1^3$ и $0<\lambda<\lambda^{*}$. b) График функции $h(z)$ при $z=x_3^3$ и $\lambda>\lambda^{*}$.}
\end{center}

\begin{center}
\includegraphics[width=6cm]{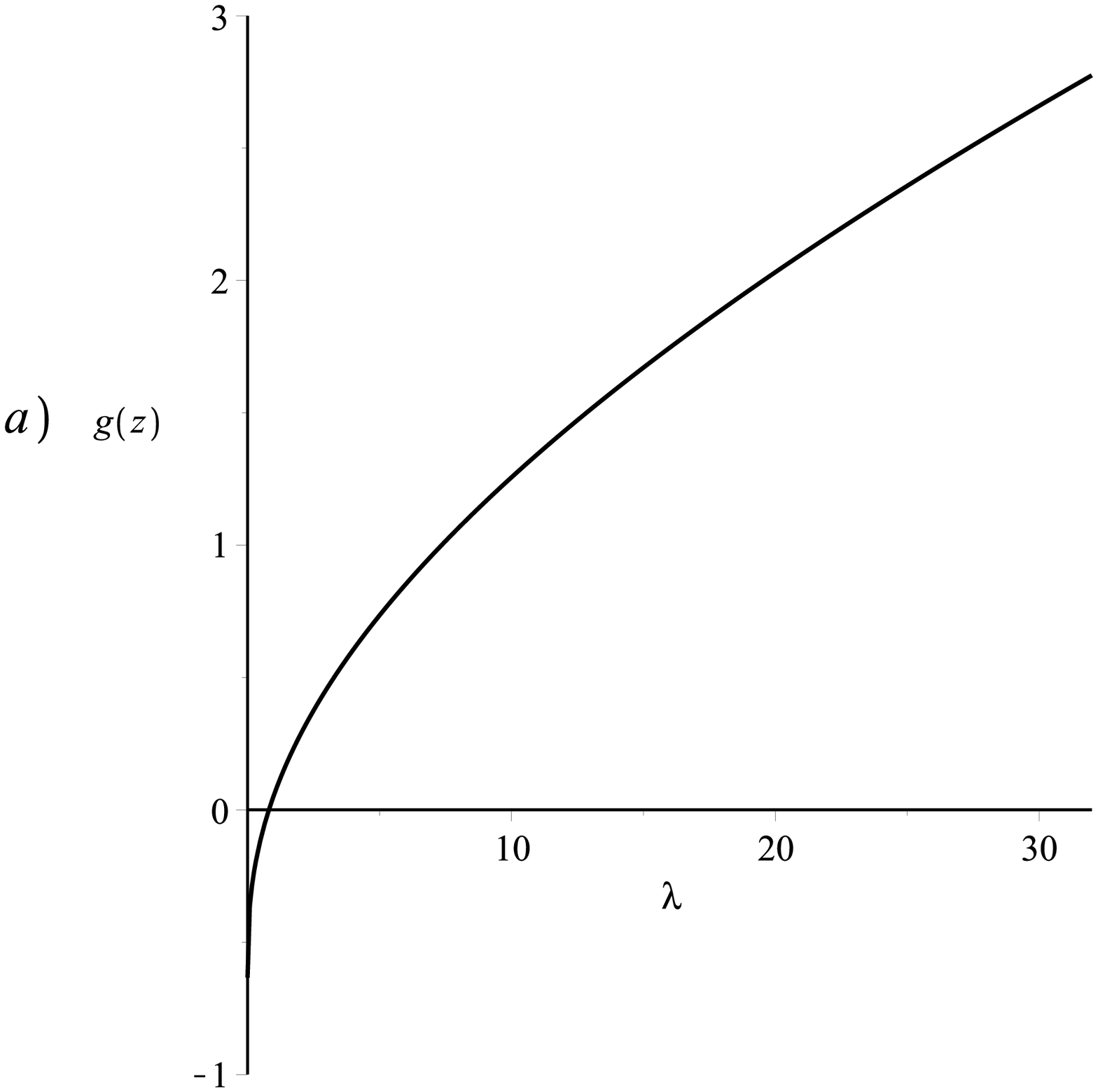} \ \includegraphics[width=6cm]{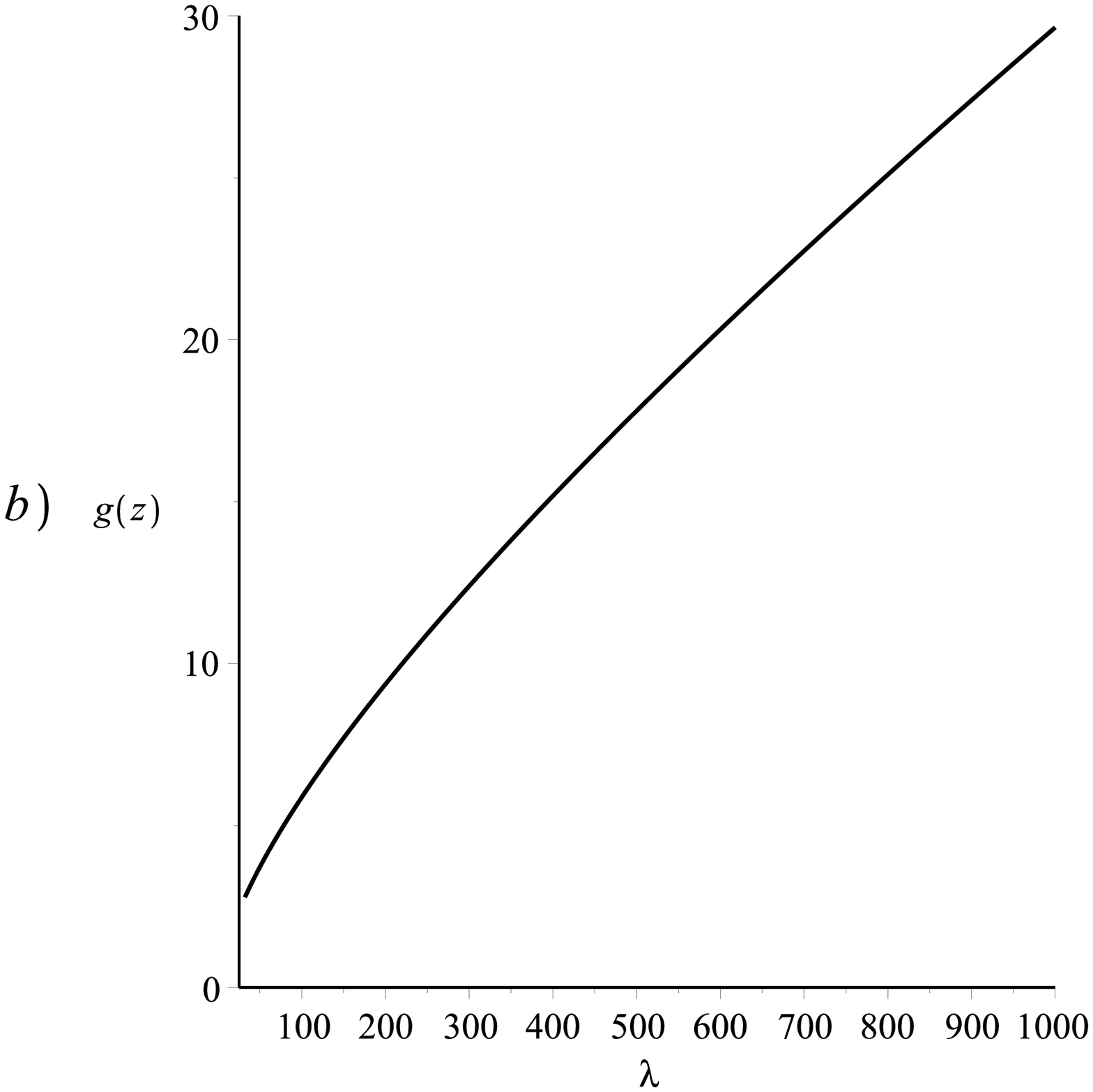}
\end{center}
\begin{center}{\footnotesize \noindent
 Рис.~2. a) График функции $g(z)$ при $z=x_1^3$ и $0<\lambda<\lambda^{*}$. b) График функции $g(z)$ при $z=x_3^3$ и $\lambda>\lambda^{*}$.}
\end{center}

\textit{Крайность меры $\mu_0$.} Приведем необходимые определения из работы \cite{MSW}.
Если удалить произвольное ребро $\langle x^0, x^1\rangle=l\in L$ из дерева Кэли $\Im^k$, то оно разбивается на две компоненты $\Im^k_{x^0}$ и $\Im^k_{x^1}$, каждая из которых называется полубесконечным деревом или полудеревом Кэли.

Рассмотрим конечное полное поддерево $\mathcal T$, которое содержит все начальные точки полудерева $\Im^k_{x^0}$. Граница $\partial \mathcal T$ поддерева $\mathcal T$ состоит из ближайших соседей его вершин, которые лежат в $\Im^k_{x^0}\setminus \mathcal T$. Мы отождествляем поддерево $\mathcal T$ с множеством его вершин. Через $E(A)$ обозначим множество всех ребер $A$ и $\partial A$.

В \cite{MSW} ключевыми являются две величины  $\kappa$ и $\gamma$. Оба являются свойствами множества мер Гиббса $\{\mu^\tau_{{\mathcal T}}\}$, где граничное условие $\tau$ фиксировано и $\mathcal T$ является произвольным, начальным, полным, конечным поддеревом $\Im^k_{x^0}$. Для данного начального поддерева $\mathcal T$ дерева $\Im^k_{x^0}$ и вершины $x\in\mathcal T$ мы будем писать $\mathcal T_x$ для (максимального) поддерева $\mathcal T$  с начальной точкой в $x$. Когда $x$ не является начальной точкой $\mathcal T$, через $\mu_{\mathcal T_x}^s$ обозначим меру Гиббса, в которой "предок"\  $x$ имеет спин $s$ и конфигурация на нижней границе ${\mathcal T}_x$ (т.е. на $\partial {\mathcal T}_x\setminus \{\mbox{предок}\ \ x\}$) задается через $\tau$.

Для двух мер $\mu_1$ и $\mu_2$ на $\Omega$ через $\|\mu_1-\mu_2\|_x$ обозначим расстояние по норме
$$\|\mu_1-\mu_2\|_x={1\over 2}\sum_{i=0}^2|\mu_1(\sigma(x)=i)-\mu_2(\sigma(x)=i)|.$$
Пусть $\eta^{x,s}$ есть конфигурация $\eta$ со спином в $x$, равным $s$.

Следуя \cite{MSW}, определим
$$\kappa\equiv \kappa(\mu)={1\over2}\max_{i,j}\sum_{l=0}^2|P_{il}-P_{jl}|;$$
$$\gamma\equiv\gamma(\mu)=\sup_{A\subset \Im^k}\max\|\mu^{\eta^{y,s}}_A-\mu^{\eta^{y,s'}}_A\|_x,$$
где максимум берется по всем граничным условиям $\eta$, всеми $y\in \partial A$, всеми соседями $x\in A$ вершины $y$ и всеми спинами $s, s'\in \{0,1,2\}$.

Достаточным условием крайности  меры Гиббса $\mu$ является $k\kappa(\mu)\gamma(\mu)<1$.

Из работы \cite{RKh1} известно, что $\kappa=\frac{z}{z+1}$ при $i\neq j$ и $\kappa=\gamma$.
Значит, для крайности меры $\mu_0$ достаточно выполнение неравенства $3\left(\frac{z}{z+1}\right)^2-1<0$ и оно верно
при $\lambda<\hat{\lambda}$.

Итак, доказана следующая теорема.

\textbf{Теорема 2.} Пусть $k=3$. Тогда для НС-модели в случае $G=\textit{петля}$ мера $\mu_0$ при $\lambda>\hat{\lambda}$
является не крайней и при $0<\lambda<\hat{\lambda}$ является крайней, где $\hat{\lambda}\approx0.8094705632$.\

\textbf{Случай $G=\textit{жезл}$}. В этом случае предполагая $z_x=z$, из (\ref{rus2.4}) получим
следующую систему уравнений:
\begin{equation}\label{rus5.1} \left\{\begin{array}{ll}
z_1=\lambda\left({1+ z_1\over z_1+z_2}\right)^k,\\[2mm]
z_2=\lambda\left({1+z_2\over z_1+z_2}\right)^k
\end{array}\right.
\end{equation}

Для ТИМГ в случае $G=\textit{жезл}$ известны следующие факты:

\begin{itemize}
\item[$\bullet$] В случае $k=2$ ($k=3$) доказано, что при $\lambda\leq1$ ($\lambda\leq\frac{4}{27}$)
существует ровно одна ТИМГ $\nu_0$, при $\lambda>1$ ($\lambda>\frac{4}{27}$) существуют ровно
три ТИМГ $\nu_0, \nu_1, \nu_2$ (см. \cite{Ro}, \cite{XR1}).

\item[$\bullet$] В случае $k>3$ доказано, что при $\lambda\leq\lambda_{cr}$ существует ровно
одна ТИМГ, при $\lambda>\lambda_{cr}$ существуют не менее трех ТИМГ, где
$\lambda_{cr}={1\over k-1}\cdot\left({2\over k}\right)^k$ (см. \cite{RKh1}).

\item[$\bullet$] В случае $k=2$ показано, что мера $\nu_0$ при $0<\lambda<\lambda_0$ и меры
$\nu_1, \nu_2$ при $1<\lambda<\lambda_1$ являются крайними и мера $\nu_0$ при $\lambda>\lambda_0$
не является крайней, где $\lambda_0\approx 2.287572$, $\lambda_1\approx 1.303094$ (см. \cite{RKh1}).

\item[$\bullet$] В случае $k=3$ доказано, что на одном из инвариантов при $\lambda\geq\frac{128}{27}$ существует ровно одна
$G^{(2)}_k$-периодическая мера Гиббса $\bar{\nu}_0$, которая является трансляционно-инвариантной и при $0<\lambda<\frac{128}{27}$ существует ровно три $G^{(2)}_k$-периодические меры Гиббса $\bar{\nu}_0, \bar{\nu}_1, \bar{\nu}_2$, где $\bar{\nu}_1, \bar{\nu}_2$ являются $G^{(2)}_k$-периодическими (не трансляционно-инвариантными).(см. \cite{KhU}).
\end{itemize}

Рассмотрим (\ref{rus5.1}) при $k=3$:
\begin{equation}\label{rus5.2} \left\{\begin{array}{ll}
z_1=\lambda\left({1+ z_1\over z_1+z_2}\right)^3,\\[2mm]
z_2=\lambda\left({1+z_2\over z_1+z_2}\right)^3.
\end{array}\right.
\end{equation}

Известно \cite{XR1}, что единственной ТИМГ $\nu_0$ соответствует единственное решение уравнения
$$z=\lambda\left(\frac{1+z}{2z}\right)^3,$$
которое получается из (\ref{rus5.2}) при $z_1=z_2=z$.
После обозначений $\sqrt[3]{z}=x,  \sqrt[3]{\lambda}=a$, из последнего уравнения получим уравнение
$2x^4-ax^3-a=0$. Аналогично случаю $G=\textit{петля}$, решив это уравнение, используя методы Феррари и Кардано из линейной алгебры,
можем получить следующее утверждение.

\textbf{Утверждение 2.} Единственное положительное решение системы уравнений (\ref{rus5.2}) при $\lambda>0$ имеет вид $(z,z)=(x^3,x^3)$,
где
\begin{equation}\label{rus5.3}
x=\frac{1}{2}\sqrt{2t_0(a)}+\frac{1}{2}\sqrt{-2t_0(a)+\frac{a^3}{64}\sqrt{\frac{2}{t_0(a)}}+\frac{3a^2}{16}}+\frac{a}{8},
\end{equation}
$a=\sqrt[3]{\lambda}$ и
$$t_0(a)=\frac{1}{24}\sqrt[3]{-108a^3+12\sqrt{81a^6+6144a^3}}-\frac{4a}{\sqrt[3]{-108a^3+12\sqrt{81a^6+6144a^3}}}+\frac{1}{32}a^2.$$

Итак, мы имеем ТИМГ $\nu_0$, соответствующую этому решению. Известно \cite{RKh1}, что матрица вероятностных переходов
для этой меры имеет вид ($z_1=z_2=z$):
\begin{equation}\label{rus5.4}
\mathbb P=\left(%
\begin{array}{cccccc}
 0 &  {1\over 2} &  {1\over 2} \\[3 mm]
  {1\over 1+ z} & { z\over 1+ z} & 0 \\[3 mm]
  {1\over 1+ z} & 0 & { z\over 1+ z} \\
  \end{array}%
\right)
\end{equation}
и ее собственные значения:
$$s_1=-{1\over z+1}, \ \ s_2={z\over z+1}, \ s_3=1.$$
Отсюда
$$s_0=max\{|s_1|, |s_2|\}=\left\{%
\begin{array}{ll}
    |s_1|, \ \ \ $ если $ {z<1} $,$ \\
   |s_2| ,\ \ \  $   если $ {z>1} $, $ \\
\end{array}%
\right.$$

\textit{Не крайность меры $\nu_0$.} Проверим условие Кестена-Стигума $ks_0^2>1$, где $s_0-$ второе максимальное по абсолютной величине
собственное значение матрицы $\mathbb P$.

Так как выражение для $z$ очень громоздкое, восьползуемся компьютерным анализом. С помощью программы MapLE
можно увидеть, что $l(z)=z-1>0$ при любых $\lambda>\dot{\lambda}\approx1$ и $l(z)=z-1<0$ при любых $\lambda<\dot{\lambda}$ (см. Рис.3).
Значит, условие не крайности при $\lambda>\dot{\lambda}$ имеет вид: $3\left({z\over z+1}\right)^2>1$, т.е. $q(z)=(\sqrt{3}-1)z-1>0.$
А при $\lambda<\dot{\lambda}$ оно имеет вид: $3\left({1\over z+1}\right)^2>1$, т.е. $w(z)=z-\sqrt{3}+1<0.$  Из неравенства $q(z)>0$ следует,
что оно верно при $\lambda>\check{\lambda}\approx2.103133692$, а из неравенства $w(z)<0$ следует, что оно верно при $\lambda<\tilde{\lambda}\approx0.4421534328$, т.е. мера $\nu_0$ не является крайней при $\lambda\in (0,\tilde{\lambda})\cup (\check{\lambda},\infty)$ (см. Рис.4).

\begin{center}
\includegraphics[width=6cm]{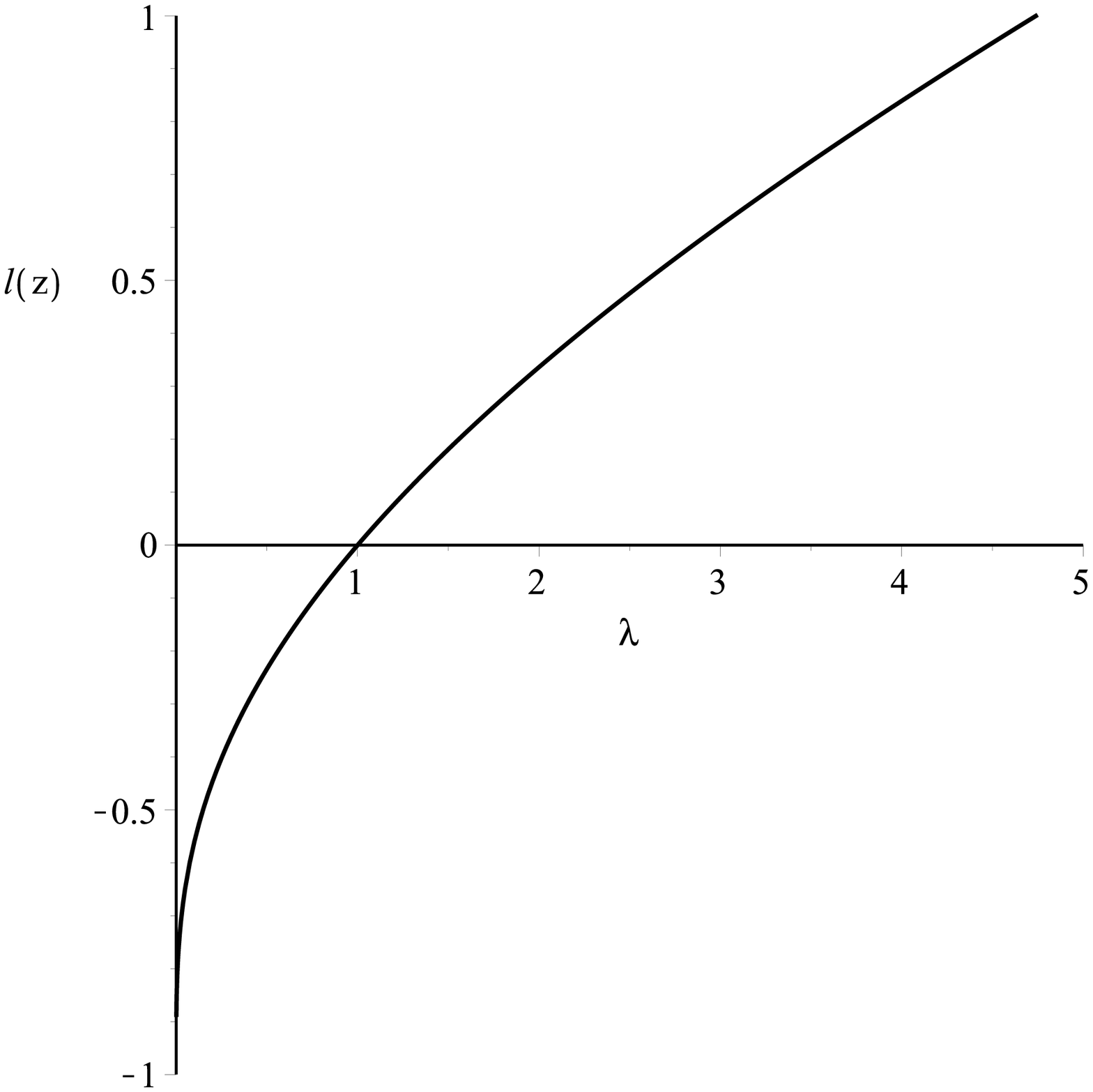}
\end{center}
\begin{center}{\footnotesize \noindent
 Рис.~3.  График функции $l(z)$ при $\lambda>0$.}
\end{center}

\begin{center}
\includegraphics[width=6cm]{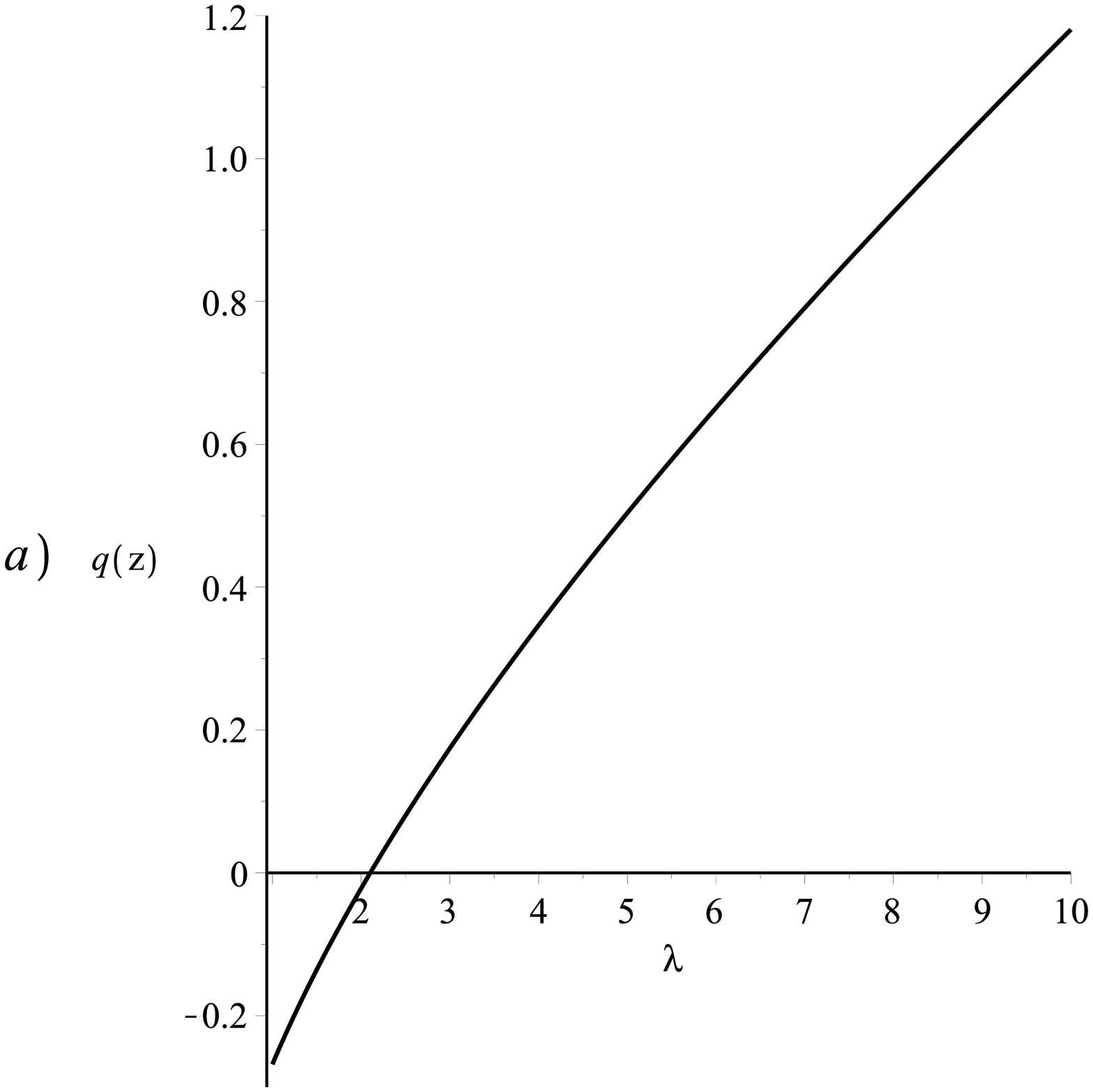} \ \includegraphics[width=6cm]{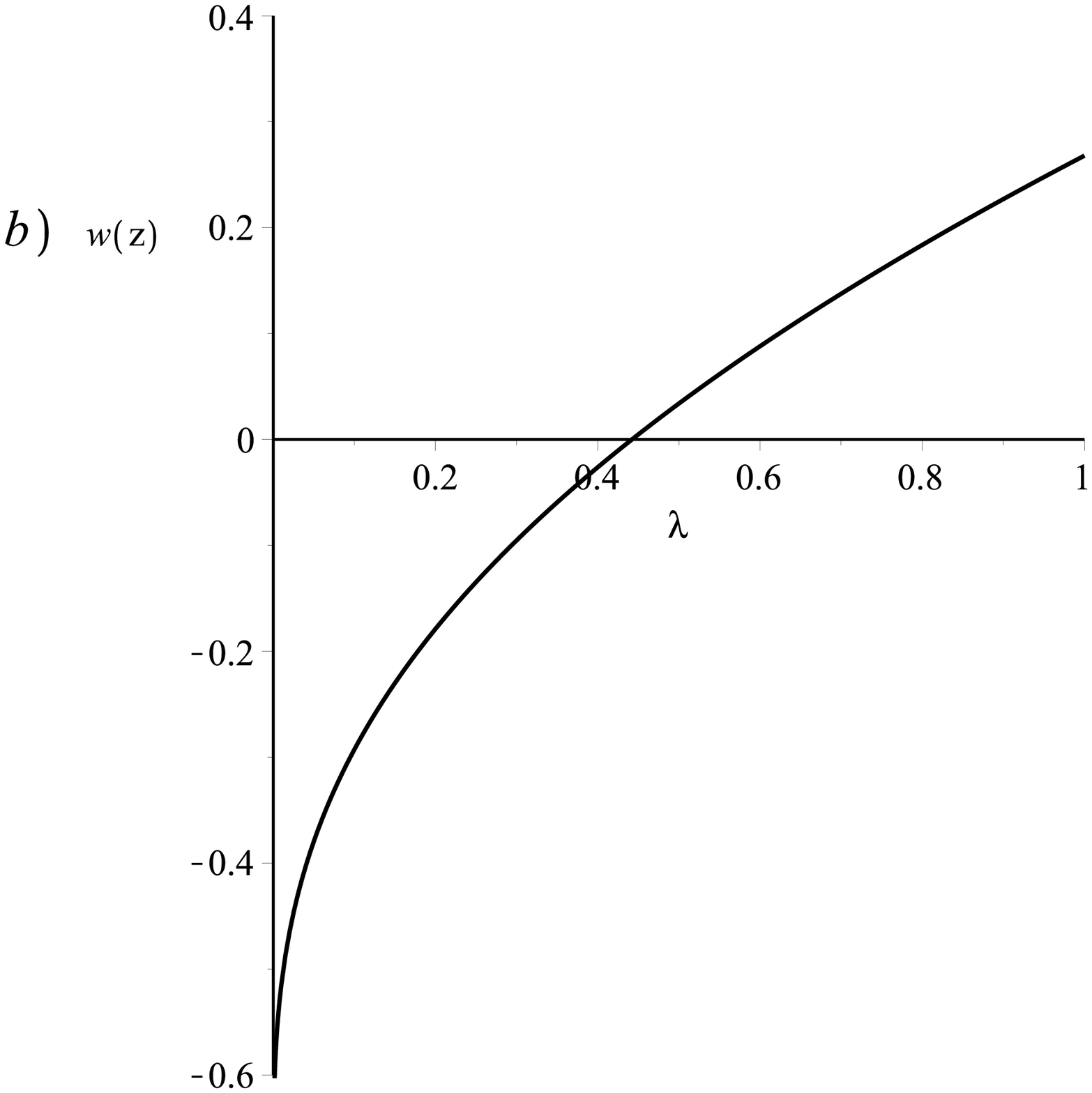}
\end{center}
\begin{center}{\footnotesize \noindent
 Рис.~4. a) График функции $q(z)$ при $\lambda>\dot{\lambda}$. b) График функции $w(z)$ при $0<\lambda<\dot{\lambda}$.}
\end{center}

\textit{Крайность меры $\nu_0$.} Из работы \cite{RKh1} известно, что при $z_1=z_2=z$
$$\kappa= \left\{%
\begin{array}{ll}
   {1\over 1+ z}, \ \ \mbox{если} \ \ {z<1},\\[2 mm]
   { z\over 1+ z}, \ \ \mbox{если} \ \ {z>1}
   \end{array}%
\right.\ $$
и $\kappa=\gamma$. Значит, для крайности меры $\nu_0$ при $z<1$ достаточно выполнение неравенства
$3\left(\frac{1}{z+1}\right)^2-1<0$, т.е. $w(z)>0$, а при $z>1$ выполнение неравенства $3\left(\frac{z}{z+1}\right)^2-1<0$,
т.е. $q(z)<0$. Первое из них верно при $\lambda>\tilde{\lambda}$, а второе при $\lambda<\check{\lambda}$ (см. Рис.4).

Таким образом, доказана следующая теорема.

\textbf{Теорема 3.} Пусть $k=3$. Тогда для НС-модели в случае $G=\textit{жезл}$ мера $\nu_0$ при $\lambda\in(\tilde{\lambda}, \check{\lambda})$
является крайней и при $\lambda\in (0,\tilde{\lambda})\cup(\check{\lambda},\infty)$ не является крайней, где $\tilde{\lambda}\approx0.4421534328$
и $\check{\lambda}\approx2.103133692$.

Справедлива следующая теорема.

\textbf{Теорема 4.} Если $k=3$, то для НС модели в случае $G=\textit{жезл}$ при $\tilde{\lambda}<\lambda <\check{\lambda}$ кроме $\nu_0$
существует по крайней мере еще одна крайняя мера Гиббса.

\textbf{Доказательство.} Пусть $k=3$. Известно, что единственная ТИМГ $\nu_0$ существует при любых значениях $\lambda>0$ и в силу
теоремы 3 она является крайней при $\tilde{\lambda}<\lambda <\check{\lambda}$. Кроме того, при $\lambda >\frac{4}{27}$ имеем меру
$\nu_0$ и две ТИМГ $\nu_1, \nu_2 $, упомянутые в вышеуказанных известных фактах в случае $G=\textit{жезл}$, а также при $0<\lambda<\frac{128}{27}$ имеем две $G^{(2)}_k$-периодические меры Гиббса $\bar{\nu}_1, \bar{\nu}_2$, отличные от трансляционно-инвариантных. Если предположим, что меры $\nu_1, \nu_2, \bar{\nu}_1, \bar{\nu}_2$ (или другие меры, если они существуют) не являются крайними в интервале $(\tilde{\lambda}, \check{\lambda})$, то остается только одна крайняя мера $\nu_0$. Но в этом случае не крайную меру нельзя выразить с помощью единственной меры $\nu_0$. Следовательно, при $\tilde{\lambda}<\lambda <\check{\lambda}$ по крайней мере одна новая мера должна быть крайней. Теорема доказана.

\textbf{Благодарности.} Авторы выражают свою глубокую признательность профессору У.А.Розикову за полезные советы.

\end{document}